\documentclass[aps,twocolumn,superscriptaddress]{revtex4}
\usepackage{graphicx}
\usepackage{amsmath}
\usepackage{amssymb}
\usepackage{bm} %bold math
\usepackage{epsf}
\usepackage[english]{babel}
\usepackage{dcolumn}
\usepackage{color}
\usepackage{epstopdf}
\usepackage{xfrac}	% to use \sfrac which make the x/y fractions of same high as text so that line spacing stays the same

\usepackage{changes}

\begin{document}

\title{Auger-induced charge migration}


%\thanks{}

%\author
%\affiliation

\author{A. Pic\'on}
\thanks{Corresponding author: antonio.picon@uam.es}
\affiliation{Departamento de Qu\'imica, Universidad Aut\'onoma de Madrid, 28049, Madrid, Spain}
\affiliation{Grupo de Investigaci\'on en Aplicaciones del L\'aser y Fot\'onica, Departamento de F\'isica Aplicada, University of Salamanca, E-37008, Salamanca, Spain}
\author{C. Bostedt}
\affiliation{Chemical Sciences and Engineering Division, Argonne National Laboratory, Argonne, Illinois 60439, USA}
\affiliation{Paul Scherrer Institute, CH-5232 Villigen, Switzerland}
\affiliation{Ecole Polytechnique FŽdŽrale de Lausanne (EPFL), CH-1015 Lausanne, Switzerland}
\author{C. Hern\'andez-Garc\'ia}
\affiliation{Grupo de Investigaci\'on en Aplicaciones del L\'aser y Fot\'onica, Departamento de F\'isica Aplicada, University of Salamanca, E-37008, Salamanca, Spain}
\author{L. Plaja}
\affiliation{Grupo de Investigaci\'on en Aplicaciones del L\'aser y Fot\'onica, Departamento de F\'isica Aplicada, University of Salamanca, E-37008, Salamanca, Spain}


\date{\today}

\begin{abstract}
Novel perspectives of controlling molecular systems have recently arisen from the possibility to generate attosecond pulses in the ultraviolet regime and tailor electron dynamics in its natural timescale. The cornerstone mechanism is the so-called charge migration, the production of a coherent charge transfer with sub-femtosecond oscillations across a molecule. Typically, charge migration is induced by the ionization of valence molecular orbitals. However, recent technological developments allow the generation of attosecond pulses in the x-ray regime. In this case, the absorption of photons creates core-hole states. In light elements, core-hole states mainly decay by Auger processes that, driven by electron correlations, involve valence orbitals. We theoretically demonstrate in a fluoroacetylene molecule a double-hole charge migration triggered by attosecond core-electron photoionization, followed by Auger electron relaxations. { This opens a new route for inducing with x rays charge transfer processes in the sub-femtosecond time scale.}
\end{abstract}


%\keywords{}

\maketitle

% References should be done using the ~\cite, \ref, and \label commands
%\section{}
%\subsection{}
%\subsubsection{}

% If in two-column mode, this environment will change to single-column
% format so that long equations can be displayed. Use
%\begin{widetext}
%\end{widetext}

% figures should be put into the text as floats.
% Use the graphics or graphicx packages (distributed with LaTeX2e)
% and the \includegraphics macro defined in those packages.
% See the LaTeX Graphics Companion by Michel Goosens, Sebastian Rahtz,
% entire page. There is no need to do explicit centering.

% \begin{figure}
% \end{figure}

% Surround figure environment with turnpage environment for landscape
% figure
% \begin{figure}
% \end{figure}

%\section{Nature Physics letter style}

%%%%%%%%%%%%%%%%%% INTRO %%%%%%%%%%%%%%%%%%%

\section{Introduction}
The capability to produce ultrashort light pulses in the attosecond (10$^{-18}$ s) time regime, allows the possibility to take snapshots of electron processes in Physics, Chemistry, and Biology. Attosecond pulse durations permit to track and study the fast dynamics of electrons –the fastest physical entities that play a major role in a chemical reaction–, unveiling key mechanisms in the microscopic scale that give rise to the understanding of the macroscopic response \cite{Plaja2013}. Those ultrashort pulses do not only serve to observe the fast motion of electrons, but they also provide the tools to tailor the electron dynamics and control matter in an unprecedented way \cite{Lepine2014,Leone2014}. A preeminent example is charge migration \cite{Nisoli2017,Worner2017}, a unique charge control only achievable with the development of attosecond light pulses.\\

Charge and energy transport plays a fundamental role in relevant chemical and biological processes. Charge migration refers to the fast motion of electrons driven purely by electron effects right after photo-excitation, occurring between hundreds of attoseconds to few femtoseconds. In first theoretical papers, charge migration was conceived as a non-equilibrium charge distribution in the molecular cation after sudden ionization of the molecule. The sudden ionization is treated as the removal of a molecular orbital and electron correlations induces the charge transfer across the molecule \cite{Cederbaum1999,Breidbach2005,Remacle2006}. Development of extreme ultraviolet (EUV) sub-femtosecond pulses with unprecedented spatio-temporal properties, and theory developments of sophisticated theoretical methods \cite{Nisoli2017}, slowly change this picture and the understanding of the control of electron motion in complex molecular structures. The broad bandwidth and coherence characteristic of such attosecond EUV pulses allow the ionization of several valence orbitals, leaving the molecule in a coherent superposition of cation states. The time evolution of the superposition induces the motion of the charge density across the molecule. In these recent years, evidences of charge migration have been experimentally demonstrated \cite{Nisoli2017,Calegari2014,Kraus2015}. Interestingly, the induced dynamics of electrons precede the nuclei motion, and one may exploit this in order to drive the nuclei response and control their motion. Indeed, charge migration holds a great promise in the control of biologically relevant molecules \cite{Calegari2014,Lara2016}.\\

%A new tendency to extend attosecond pulses into the x-ray regime has grown in the recent years.
Attosecond pulses in the EUV regime are by now routinely generated by focusing strong-field infrared (IR) lasers on atomic gas phase targets, resorting to the extreme non-linear process known as high-harmonic generation. Further technical developments not only allow to increase the flux of the generated attosecond pulses, but also to increase the photon energies up to the soft x-ray region \cite{Popmintchev2012,Silva2015}. These pulses present a formidable spatio-temporal coherence and allow to perform time-resolved studies with the characteristic atomic resolution of x rays and the characteristic temporal resolution of attosecond pulses. New opportunities arise from theoretical proposals to produce attosecond pulses at free-electron lasers (FELs) \cite{Zholents2005,Prat2015}. Very recently  attosecond coherent pulses in the hard x-ray regime were demonstrated at the Linac Coherent Light Source \cite{Huang2017}. Experimental implementations are planned at the Linac Coherent Light Source and SwissFEL. The high flux of FEL pulses provide the perspective to use the broad bandwidth of these pulses for non-linear spectroscopy techniques. \\

%%%%%%% Figure
\begin{figure}[t]
\includegraphics[trim=0cm 0cm 0cm 0cm,clip,width=8.cm]{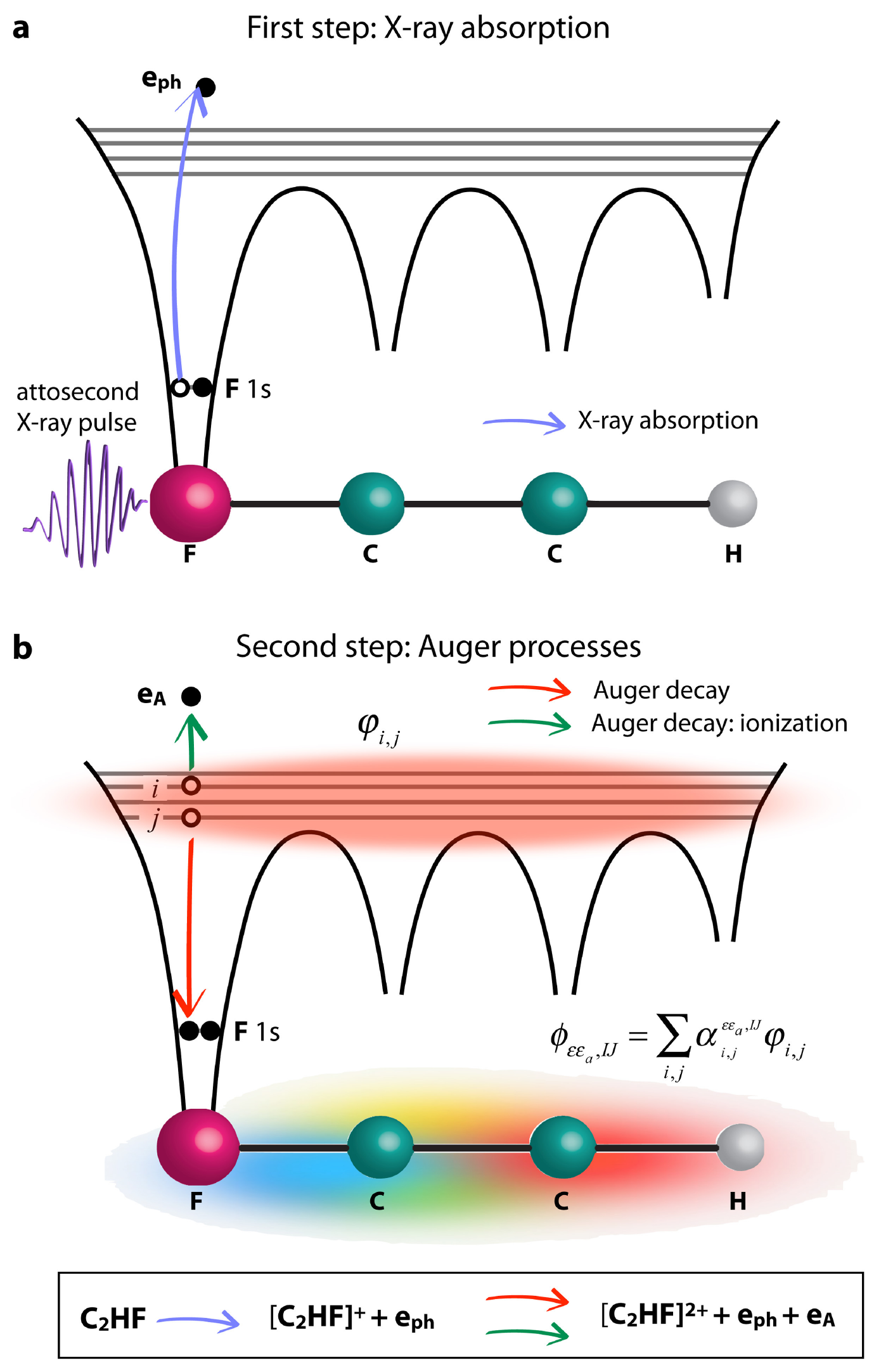}
\caption{{\bf X-ray ionization and Auger electron relaxation}. Scheme of the physical scenario. {\bf a}, First step, the attosecond x-ray pulse ionizes the F 1s orbital, creating a photoelectron (e$_{\rm ph}$) in the continuum. {\bf b}, Second step, core-hole mainly decays via Auger processes, involving two electrons over the core-hole orbital. One electron fills the core vacancy and releases energy that is transferred to the second electron, which is ionized producing an electron in the continuum known as Auger electron (e$_{\rm A}$). After the Auger processes, the molecule remains in a superposition of dication states inducing a charge migration in the valence shell.}
\label{fig1}
\end{figure}

{ These developments have triggered studies of charge migration, especially using nonlinear schemes with x rays \cite{Biggs2013,Mukamel2013}, but also via photoionization \cite{Kuleff2016}}. While IR to EUV photons mainly interact with valence electrons, x-ray photons strongly interact with core electrons. This is an important difference, as core orbitals are well-localized around atomic sites. The x-ray ionization promotes those electrons into the continuum, leaving behind a deep vacancy, or core-hole state. Although this first excitation is local, the subsequent decay of the core-hole state may involve delocalized valence orbitals. In light atoms, which are the mainly ones forming biological molecules, the decay of core-hole states are dominated by Auger processes \cite{Bambynek1972}. Auger transitions are originated by strong electron correlations. Characteristic core-hole lifetimes span from hundreds of attoseconds to a few of femtoseconds. Hence, x-ray interactions may be considered also as effective valence double-hole excitations via Auger processes. { Note that this is a different mechanism than previous works in double-hole charge migration, in which the double-hole distribution is created via valence excitation \cite{Hollstein2017}. It is, therefore, timely to investigate the  possibility of inducing charge migration via core-electron ionization.} \\

In this work we present a theoretical study of auger-induced charge migration (AICM) in fluoroacetylene (C$_2$HF) driven by attosecond soft-x-ray pulses. Monohaloacetylenes have importance in atmospheric chemistry, global-warming-potential refrigerants, and combustion applications \cite{Khiri2016,McLinden2017}. C$_2$HF is a linear molecule with a fluorine atom in at one end, two carbon atoms at the middle, and a hydrogen atom on the other end, see Fig. \ref{fig1}.  An attosecond pulse in the soft-x-ray regime of 800 eV can effectively ionize the 1s electron from the fluorine atom. Because this is the dominant excitation, the ionization from other shells can be neglected. A coherent charge migration is induced across the molecule, with a prominent effect on the hydrogen atom located at the other side of the excitation as will be shown below. The theoretical model developed for AICM is general and can also be extended to more complex molecules, taking full advantage of the novel x-ray sources producing attosecond pulses. This work opens the door to future studies aiming at controlling matter in a unique way by combining coherence with the characteristic electron correlations of core-hole states.\\

\section{Results}

\subsection{Auger-induced charge migration in C$_2$HF}

We consider the fluoroacetylene molecule interacting with an 800-eV attosecond pulse whose envelope is modelled as a Gaussian function with a full-width at half-maximum (FWHM) of 160 as. { The pulse is linearly polarized along the molecular axis}. At this photon energy, the ionization of the 1s electron from the F atom is the dominant channel compared to the ionization from other shells, therefore, for the sake of simplicity, we consider only this channel. The ionization potential of the 1s F electron is calculated to be 717 eV, hence, in a first step, the x-ray pulse generates photoelectrons with a central energy around 83 eV, see Fig. \ref{fig1}. The molecule remains in a cation state with a core hole that, in a second step, decays mainly by Auger processes. The Auger transitions involve two electrons in orbitals above the core hole, among them valence orbitals. One electron fills the core vacancy transferring its energy to the second electron, which is excited to the continuum as an Auger electron. The final product is, thus, a dication state with two electrons in the continuum: a photoelectron and an Auger electron. There are 190 possible dication states, in which the largest Auger transitions correspond to singlet states. In this work we restrict then to the 55 final singlet states. The dynamics can be described by \\

\begin{eqnarray} 
\psi (t) = b_0 ({\bf R},t) \Phi_{0} ({\bf X},{\bf R})  + \sum_\varepsilon \sum_i b_{\varepsilon ; i}({\bf R},t) \Phi_{\varepsilon; i} ({\bf X},{\bf R}) \nonumber \\
+ \sum_{\varepsilon \varepsilon_a} \sum_{IJ}  b_{\varepsilon \varepsilon_a ; IJ}({\bf R},t) \Phi_{\varepsilon \varepsilon_a ; IJ} ({\bf X},{\bf R}) \; , \nonumber \\ \label{ansatz_Auger_R}
\end{eqnarray}
in which $\Phi_{0} ({\bf X},{\bf R})$ refers to the ground state of the molecule, $\Phi_{\varepsilon; i} ({\bf X},{\bf R})$ refers to the core-hole state together with a photoelectron state, and $\Phi_{\varepsilon \varepsilon_a ; IJ} ({\bf X},{\bf R})$ refers to the final dication state together with the photoelectron and the Auger electron states. The continuum states are expanded in partial waves and they are defined by the energy, the spin, and the two quantum numbers related to the angular momentum. The dication state is described as a multiconfigurational superposition of  Hartree-Fock orbitals (see more details in { appendix B}). The dynamics of the system are calculated by solving the time-dependent Schr\"odinger equation as described in Ref. \cite{Picon2017A}.

%%%%%%% Figure
\begin{figure*}[t]
\includegraphics[trim=0cm 0cm 0cm 0cm,clip,width=18.cm]{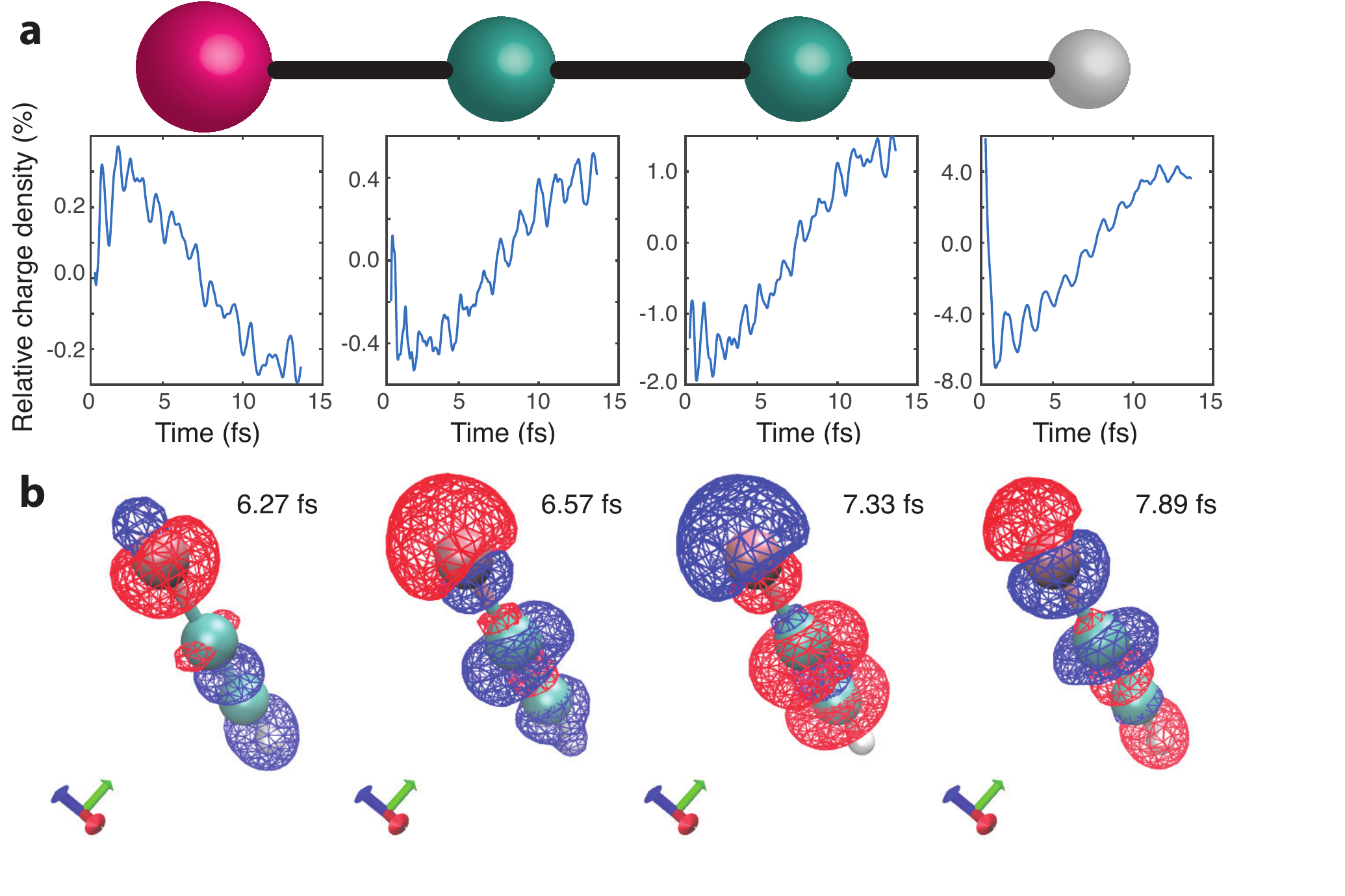}
\caption{{\bf Coherent charge migration}. {\bf a}, The percentage of the hole charge density difference $\Delta n_{2h}(t)$ is calculated around each atom of the C$_2$HF molecule. The changes in the electron density are due to the coherent part, off-diagonal terms, of the density matrix (\ref{reduced_matrix}). The coherent part presents very fast oscillations. {\bf b}, Coherent part of the charge density $n_{2h}(t)$ for four different times. Blue isosurface is related to depletion of charge, while red isosurface is related to gain of charge. The charge oscillation from positive to negative in the hydrogen atom occurs in less than 1.5 fs.
}
\label{fig2}
\end{figure*}

The charge migration originates by the coherent superposition of dication states. During the dynamics of the system, we need to consider then the two continuum electrons entangled with the dication states. In order to calculate the charge density evolution of the dication state, we take the reduced density matrix of the system by integrating over the continuum states

\begin{eqnarray}
\rho^{(N-2)} (t)= \sum_{IJ} \sum_{I'J'} \sum_{\varepsilon \varepsilon_a} b_{\varepsilon \varepsilon_a ; IJ}(t) b_{\varepsilon \varepsilon_a ; I'J'}^*(t) \times  \,\,\,\,\,\,\, \nonumber \\
 \vert \{ \Phi_{IJ}^{(N-2)} \} \rangle  \langle \{ \Phi_{I'J'}^{(N-2)} \} \vert \;,
\nonumber \\ \label{reduced_matrix}
\end{eqnarray}

The off-diagonal terms of Eq. (\ref{reduced_matrix}) are related to the coherent interferences between different dication channels. These terms give rise to the induced charge migration across the molecule as we detail below. { Note that we are integrating over all quantum numbers of the continuum states, including the angular momenta, assuming that no detection of outgoing electrons is performed}. From the reduced density matrix of the dication state (\ref{reduced_matrix}) the charge density is calculated by integrating over all electron coordinate variables except one \\
\begin{eqnarray}
n_d({\bf x}_1) = \int d{\bf x}_2^3 \int d{\bf x}_3^3 ... \int d{\bf x}_{N-2}^3   \times \hspace{1cm} \nonumber \\
\langle {\bf x}_1 {\bf x}_2... {\bf x}_{N-2} \vert \rho^{(N-2)} \vert {\bf x}_1 {\bf x}_2... {\bf x}_{N-2} \rangle \;, \label{dication_charge_den}
\end{eqnarray}
and the time-dependent two-hole-valence charge distribution is described then by the difference of the ground state and dication charge density $n_{2h}(t)=n_{GS}-n_d(t)$. Naturally, if we do not consider the cross terms of the density matrix (\ref{reduced_matrix}), we obtain a two-hole charge density $n_{inc}(t)$ with no interferences among different states. We measure the changes due to the coherent part by taking the normalized difference  $\Delta n_{2h}(t)=(n_{2h}(t)-n_{inc}(t))/n_{inc}(t)$. \\

The two-hole charge density is calculated for C$_2$HF after the x-ray ionization of the F 1s orbital, see Fig. \ref{fig2}. Once the charge density is calculated, we can obtain the local charge density around each atom in time. The incoherent part, $n_{inc}(t)$, gives rise to the expected exponential increase of charge distribution, but this charge do not present a motion across the molecule. The largest Auger transitions involve molecular orbitals partially located at the F site, therefore the density of valence holes is mainly located around the F atom. However, Auger transitions also affect distant atoms such as the hydrogen atom. Now we focus on the coherent part of the two-hole charge density $n_{2h}(t)$, whose contribution is represented in Fig. \ref{fig2}a. We observe two type of dynamics: a slow and a fast oscillation. The slow oscillation brings charge from the F atom to the H atom in around 15 fs, while the fast oscillations present subfemtosecond dynamics across the molecule, see Fig. \ref{fig2}b. These subfemtosecond oscillations result from the electron dynamics, since the molecular ions remain fixed. Interestingly, this Auger-induced charge migration extends from one end of the molecule to the other. { This naturally opens questions about the dependencies of these oscillations, for example in a distant atom as the hydrogen atom, by tailoring the parameters of the attosecond pulse}.  \\

\subsection{ Attosecond pulse effects on AICM}

\begin{figure*}[t]
\includegraphics[trim=0cm 0cm 0cm 0cm,clip,width=12.cm]{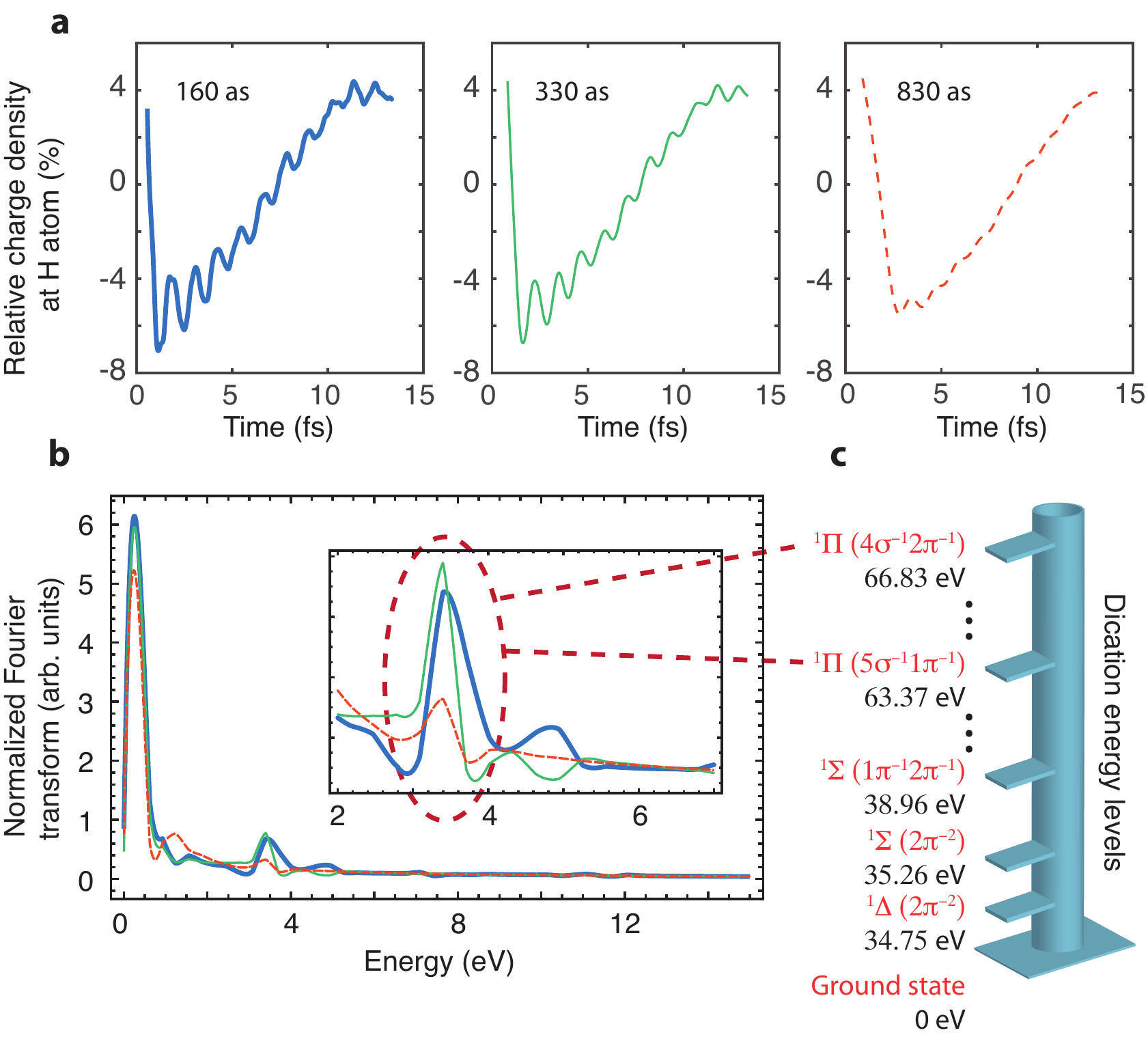}
\caption{{ \bf Pulse dependence on the auger-induced charge migration}. {\bf a}, Calculated AICM around the hydrogen atom of the C$_2$HF molecule for a pulse length of 160 as (the blue thick line), 330 as (the green thin line), and 830 as (the red dashed line) FWHM. The charge local density is strongly dependent on the pulse duration of the attosecond pulse. {\bf b}, Fourier transform of the charge oscillations for the three different pulses. The frequency approximately at 9 eV photon energy decreases with the increase of the pulse length. This high frequency component is originated from a superposition achievable only with a broad bandwidth. {\bf c}, From all final dication states, a couple of degenerate states contribute to the high frequency oscillation of AICM.  }
\label{fig3}
\end{figure*}

{
AICM produces a fast electron dynamics across the molecule. We shall now see that, by changing the pulse length of the attosecond pulse, we are able to manipulate the charge oscillation at the distant H atom. Fig. \ref{fig3}a shows the charge density oscillations at the hydrogen ion site, triggered by photonionization from an attosecond pulse with different durations. For the shortest pulse length, the charge density shows a distinctive fast oscillation while, as the pulse length increases, the time profile of the oscillation changes significantly. For pulses longer than 2-fs FWHM these oscillations are considerably small. A unique dynamics is then achieved by coherent x-ray ultrashort pulses with pulse durations competing with the characteristic times of Auger electron relaxations. This represents a complete new regime to be explored with the advent of novel x-ray sources.\\
}
In Fig. \ref{fig3}b, we show the Fourier analysis of the charge density oscillations. There are several peaks at different frequencies, up to approximately 5 eV photon energy, whose amplitudes are sensitive to the pulse duration. Each of the peaks describe the oscillation induced by a superposition of dication states, separated by the corresponding energy. Fig. \ref{fig3}c shows the particular case of the high peak around 3.4 eV. The relative intensity of the different Fourier components measures the degree of population of each dication state superposition. It can be observed that the number of states in the excited superposition excited decreases with the length of the attosecond pulse. In order to understand the underlying mechanism of AICM, we should think about the response of the molecule to the x-ray interaction within a one-step model \cite{Aberg1982,Aberg1992} that can be extended to a time-dependent framework \cite{Sullivan2016}, in which the final dication state excitation depends on the  amount of energy pumped into the system. { See appendix A for more details about the underlying mechanism}. A pulse of 160-as FWHM has a bandwidth of approximately 11-eV FWHM, enabling the excitation of a broad coherent superposition, similarly to standard charge migration that is induced by the ionization from a broad-bandwidth pulse. The promising perspectives of producing and tailoring charge migration in biomolecules now can be extended to the x-ray regime. The fact of having access to a double-hole excitation in the valence, and to a larger bandwidth at such photon energies, introduces an attractive knob to tailor the electron dynamics. Novel capabilities to perform hetero-site pump-probe studies \cite{Picon2016B}, consisting in the use of a pump and a probe x-ray pulse with different photon energy, are ideal to induce and observe AICM in a molecular system. Also, by using the so-called high-harmonic spectroscopy, the charge migration can be identified by the high-harmonic spectrum generated by an IR pulse \cite{Kraus2015}. { Another possible scheme is by measuring the coherent radiation emitted by the oscillating charge, as described in Ref. \cite{Kuleff2011}. Considering feasible parameters for a free-electron laser experiment; beam waist at focus of 100 $\mu$m, enough intensity to saturate the sample, and a large sample target of around 1 cm$^3$, we obtain more than 4000 photons/fs at 3.4 eV, a signal that is the signature of the induced charge migration.}\\

%%%%%%% Figure

%%%%%%%%%%%%%%%%%%%%%%%%% DISCUSSION %%%%%%%%%%%%%%%%%%%%%%%%%%%
\section{Discussion}

In conclusion, double-hole charge migration on the order of hundreds of attoseconds can be induced in molecular systems by the use of coherent ultrashort x-ray pulses, with pulse lengths shorter than the characteristic lifetimes of core-hole states. The charge migration is purely driven by electron interactions, which creates a coherent superposition of dication states via Auger processes. { The superposition depends on the bandwidth of the pulse}. Here, we demonstrate this effect in a C$_2$HF molecule. A 800-eV 160-as FWHM pulse is used to ionize the 1s electron on the F atom, producing charge migration on the excited dication state molecule across the molecule in a sub-femtosecond time scale. This unique way for molecular control can be explored in the future with novel x-ray sources delivering coherent attosecond x-ray pulses, opening a promising perspective to extend charge migration to the x-ray regime. Developments of sub-attosecond x-ray sources \cite{Hergar2013} could even exploit this scheme further exciting a broader superposition of states. { This work significantly contributes to the aim of tailoring the response of the molecular nuclei and chemical reaction by resorting to a fast induced electron dynamics in the valence.}

%%%%%%%%%%%%%%%%%%%%%%%%% METHODS %%%%%%%%%%%%%%%%%%%%%%%%%%%
%\section{Methods}

{
\section*{Appendix A: Underlying mechanism}
The coherent oscillations in the charge density rise from the coherences between different final dication states. The final dication states are entangled with the electrons in the continuum, those are an important source of decoherence. This is reflected when we trace out the continuum part in order to obtain the reduced density matrix shown in Eq. (\ref{reduced_matrix}). The coherences are only possible for different dication states whose continuum electrons are in the same state, as we show in the scheme depicted in Fig. \ref{fig4}a. In the shown scheme the bandwidth of the pulse plays an important role, i.e. it determines the distance in energy between the dication states. We calculate the Auger spectrum in coincidence with the photoelectron, the so-called photoelectron-Auger spectrum. For a photoelectron energy of 82 eV, we obtain the Auger spectrum represented in Fig. \ref{fig4}b, expanding only in the region for the peaks corresponding to the final states $4\sigma^{-1}2\pi^{-1}$ and $5\sigma^{-1}1\pi^{-1}$. Note that the peaks overlap for the case of a short FWHM-160-as pulse, showing the possibility to have a superposition of dication states with the same continuum states.

%%%%%%% Figure
\begin{figure}[t]
\includegraphics[trim=0cm 0cm 0cm 0cm,clip,width=6.cm]{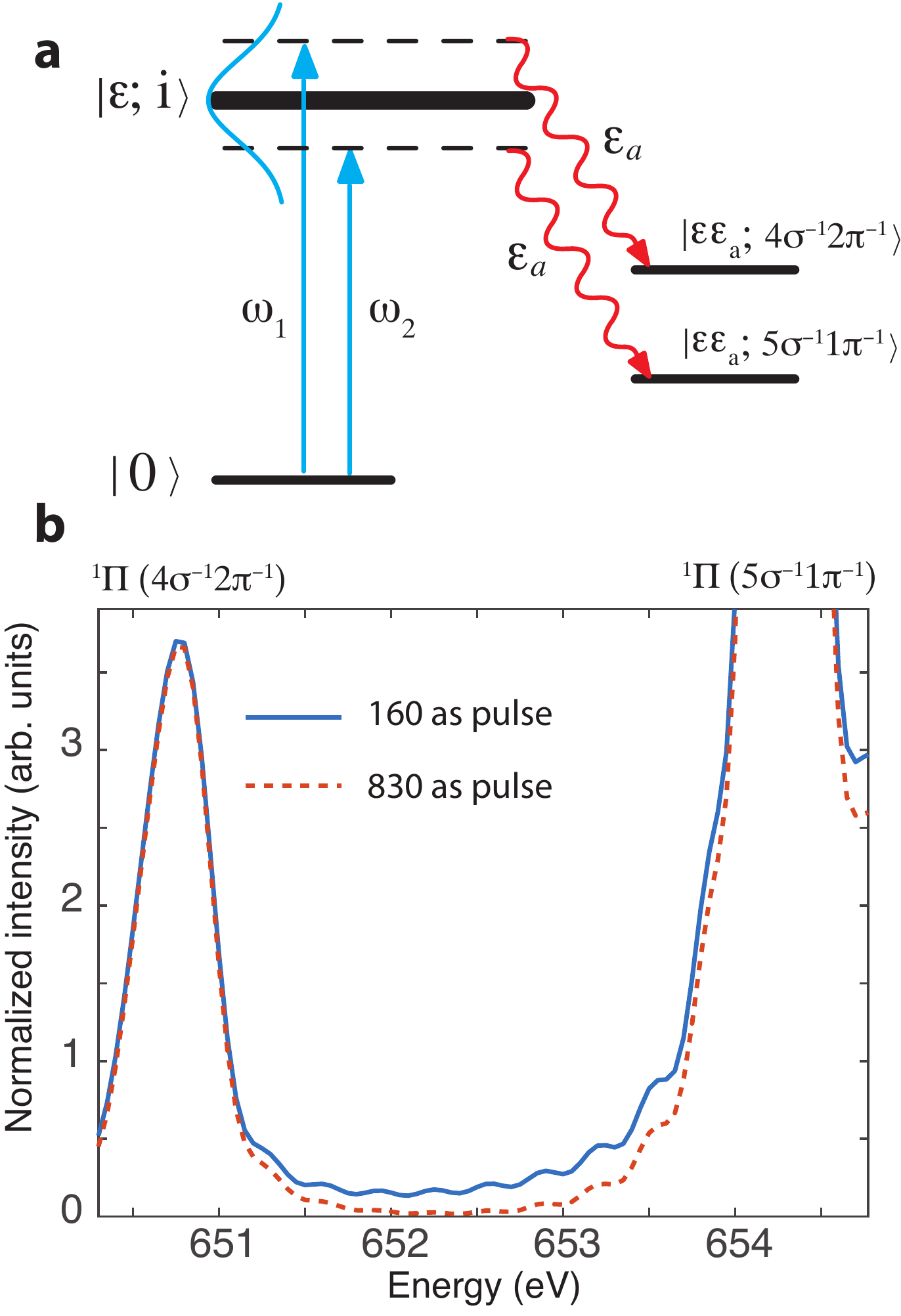}
\caption{ {\bf Interferences paths to different dication states}. {\bf a}, For an excitation of a core-excited state with a photoelectron in the state $\varepsilon$, a broadband band may populate two different dication states via absorption of photon with frequency $\omega_1$ and $\omega_2$, resulting with a final Auger electron with the same energy. {\bf b}, Auger spectrum in coincidence with a photoelectron with 82-eV energy for two different pulse lengths, FWHM 160 as and 830 as. While there is no overlapping between the Auger peaks corresponding to the final states $4\sigma^{-1}2\pi^{-1}$ and $5\sigma^{-1}1\pi^{-1}$ for the long pulse, there is a small overlapping for the short pulse.}
\label{fig4}
\end{figure}

In general, the lineshape of the Auger spectrum in coincidence with the photoelectron is given by multiplying the bandwidth of the pulse with a Lorentzian profile, whose width is determined by the core-hole lifetime. This provides a good estimate of the relevant parameters for inducing a fast electron dynamics. First, it is desirable to have a fast core-hole decay, but it is also important to have an x-ray excitation that is shorter than the characteristic decay time.

}

\section*{Appendix B: Theoretical model}
Standard charge migration is induced by the interaction of a broad bandwidth attosecond EUV pulse with a molecule. The pulse ionization produces a valence hole that migrates along the molecule. Due to the coherent character of the process, we need to simulate the dynamics both of the cation molecule and the photoelectron \cite{Lara2016}. In AICM, we need to solve time-dependent equations of motion with the Ansatz given by Eq. (\ref{ansatz_Auger_R}) involving doubly-continuum states. $\Phi_{0} ({\bf X},{\bf R})$ refers to the ground state of the molecule, $\Phi_{\varepsilon; i} ({\bf X},{\bf R})$ refers to the core-hole state together with a photoelectron state, and $\Phi_{\varepsilon \varepsilon_a ; IJ} ({\bf X},{\bf R})$ refers to the final dication state together with the photoelectron and the Auger electron states. ${\bf X}$ refers to the electron configuration of the molecule, while ${\bf R}$ refers to the nuclear degrees of freedom. Here we assume a fix nuclear geometry. Continuum orbitals are expanded in partial waves from a single center. Those are calculated using a K-matrix approach \cite{Demekhin2011} and electron correlations in the Auger and dipole transitions are considered as we detail below \cite{Kelly1975}. Continuum photoelectron states are characterized by $\varepsilon$, referring to the group of quantum numbers; spin, angular momenta of the partial wave, and energy. Similarly, Auger electron states are characterized by $\varepsilon_a$, referring to the same group of quantum numbers. Core-hole states are described at the configuration-interaction singles
\begin{eqnarray*}
\Phi_{\varepsilon ; i} ({\bf X},{\bf R}) = \Phi_{(\varepsilon L M); i} ({\bf X},{\bf R})  = \hspace{1cm} \\
{1\over\sqrt{2}} (a^\dagger_{\varepsilon L M \alpha} a_{i\alpha} + a^\dagger_{\varepsilon L M \beta} a_{i\beta}) \vert \alpha \rangle
\end{eqnarray*}
where  $\vert \alpha \rangle$ refers to the ground state at the Hartree-Fock level, $a^\dagger$ and $a$ are annihilation and creation molecular orbital operators, in which $LM$ is the angular momenta of the partial wave, $\alpha$ and $\beta$ are the spin states, and $i$ is the core orbital with the vacancy. Molecular orbitals are calculated using the PSI4 code \cite{Parrish2017}. The final states are described as
\begin{eqnarray*}
\Phi_{\varepsilon \varepsilon_a ; IJ} ({\bf X},{\bf R}) = \Phi_{(\varepsilon L M) (\varepsilon_a L_a M_a) ; IJ} ({\bf X},{\bf R}) = \hspace{2cm}  \\
{1\over\sqrt{2}} \left[ a^\dagger_{\varepsilon L M \alpha} a^\dagger_{\varepsilon_a L_a M_a \beta} - a^\dagger_{\varepsilon L M \beta} a^\dagger_{\varepsilon_a L_a M_a \alpha} \right] [ \sum_{ij} C_{IJ,ij} \, a_{i} \, a_{j} ] \vert \alpha \rangle
\end{eqnarray*}
in which the coefficients $C_{IJ,ij}$ describing the excited dication states are found at the 2-hole configuration interaction (2hCI) \cite{Agren1992} level. The time-dependent Schr\"odinger equation (TDSE) obtained from Eq. (\ref{ansatz_Auger_R}) is solved using the formalism derived in Ref. \cite{Picon2017A,Lehmann2016}. The TDSE is solved with a parallelized Runge-Kutta method.

A further improvement of the current model could account for the orbital relaxation of the core-hole state and additional excitations of the dication molecule, satellites states \cite{Puglisi2018}. As a first-order correction, it is expected that the Auger transitions would be different, modifying then the induced charge migration. Satellites could add a richer dynamics on the charge migration that has not been explored in this manuscript.

\section*{Appendix C: Charge density of two-valence-holes in  dication excited states}
After Auger processes take place, two electrons already left the molecule and we need to study the induced charge density on the dication molecule. Our dication state will be described by the reduced density matrix after tracing out the continuum orbitals of the system, obtaining Eq. (\ref{reduced_matrix}). The reduced density matrix can be expanded in Hartree-Fock molecular orbitals by using the 2hCI coefficients and the charge density (\ref{dication_charge_den}) is then written as
\begin{eqnarray*}
n_d({\bf x}_1) = \sum_{ij} B_{ij,ij} \sum_{m\neq i,j} \vert \Phi_m ({\bf x}_1) \vert^{2} + \hspace{1cm} \\
\sum_{ij[m]}\sum_{i'j'[m']\neq ij[m]} B_{ij,i'j'} \Phi_{m'} ({\bf x}_1) \Phi_{m}^*({\bf x}_1)
\end{eqnarray*}
in which the label $m$ stands for the hole (either in the $i$ or $j$ orbital) that is different, and the function $B_{ij,i'j'}$ is defined as
\begin{eqnarray*}
B_{ij,i'j'}(t) =\sum_{IJ} \sum_{I'J'}  C_{IJ,ij} C_{I'J',i'j'}^* \sum_{\varepsilon \varepsilon_a} b_{\varepsilon \varepsilon_a ; IJ}(t) b_{\varepsilon \varepsilon_a ; I'J'}^*(t)
\end{eqnarray*}
The calculation of the dication amplitudes $b_{\varepsilon \varepsilon_a ; IJ}(t)$ provides the necessary information to calculate the charge density in time.

%%%%%%%%%%%%%%%%%%%%%%%%% ACKNOWLEDGEMENTS %%%%%%%%%%%%%%%%%%%%%%%%%%%
\section{ACKNOWLEDGEMENTS}
A.P. is thankful to Alicia Palacios for very fruitful discussions and to Hans {\AA}gren who provided essential input to this work. This project has received funding from the European Union's Horizon 2020 research and innovation programme under the Marie Sklodowska-Curie Grant Agreement No. 702565, from Comunidad de Madrid through the TALENTO program with ref. 2017-T1/IND-5432, and from the U.S. Department of Energy, Office of Science, Basic Energy Sciences, Division of Chemical Sciences, Geosciences, and Biosciences through Argonne National Laboratory under contract no. DE-AC02-06CH11357. We acknowledge support from Junta de Castilla y Le\'on (Project SA046U16) and MINECO (FIS2016-75652-P). C.H.-G. acknowledges support by a 2017 Leonardo Grant for Researchers and Cultural Creators, BBVA Foundation.

\section{AUTHOR CONTRIBUTIONS}

A.P. conceived the idea with contributions from C.B. and L. P.. %  
A.P. developed the theoretical model. %
A.P., C.B., L.P., C.H.G. discussed the results and wrote the manuscript with contributions from all authors. %

{Competing financial interests: The authors declare no competing financial interests.}

\end{document}